\documentclass[preprint,epsfig,showpacs,preprintnumbers,amsmath,amssymb]{revtex4}
\usepackage{graphicx}
\begin{document}
\preprint{}
\title{QUANTAL EFFECTS ON SPINODAL INSTABILITIES IN CHARGE ASYMMETRIC NUCLEAR MATTER}

\author{S. Ayik$^{1}$}\email{ayik@tntech.edu}
\author{N. Er$^{2}$}
\author{O. Yilmaz$^{2}$}
\author{A. Gokalp$^{2}$}
\affiliation{$^{1}$Physics Department, Tennessee Technological University, Cookeville, TN 38505, USA \\
$^{2}$Physics Department, Middle East Technical University, 06531
Ankara, Turkey}

\date{\today}

\begin{abstract}
Quantal effects on growth of spinodal instabilities in charge
asymmetric nuclear matter are investigated in the framework of a
stochastic mean field approach. Due to quantal effects, in both
symmetric and asymmetric matter, dominant unstable modes shift
towards longer wavelengths and modes with wave numbers larger than
the Fermi momentum are strongly suppressed. As a result of quantum
statistical effects, in particular at lower temperatures,
magnitude of density fluctuations grows larger than those
calculated in semi-classical approximation.

\end{abstract}


\maketitle

\section{Introduction}

In many processes, such as induced fission, heavy-ion fusion near
barrier energies and spinodal instabilities and nuclear
multi-fragmentation, dynamics of density fluctuations play a
dominant role. For description of these processes mean-field
transport models, such as time-dependent Hartree-Fock (TDHF)
\cite{R1,R2} and the Boltzmann-Uhling-Uhlenbeck (BUU)
\cite{R3} models are not very useful. TDHF includes, the so called,
one-body dissipation mechanism, but associated fluctuation
mechanism is not incorporated into the model. Similarly, the
extended TDHF and its semi-classical approximation BUU model
involves one-body and collisional dissipation, but the associated
fluctuation mechanisms are not included into the description. It
is well known that no dissipation takes place without
fluctuations. In order to describe dynamics of density
fluctuations, we need to develop stochastic transport models by
incorporating fluctuation mechanisms into the description.  There
are two different mechanisms for density fluctuations: (i)
collisional fluctuations generated by two-body collisions and (ii)
one-body mechanism or mean-field fluctuations. Much effort has
been given to improve the transport description by incorporating
two-body dissipation and fluctuation mechanisms. The resultant
stochastic transport theory, known as Boltzmann-Langevin model
\cite{R4,R6}, provides a suitable framework for dynamics of
density fluctuations in nuclear collisions around Fermi energy.
However, two-body dissipation and fluctuation mechanisms do not
play an important role at low energies. At low bombarding
energies, mean-field fluctuations provide the dominant mechanism
for fluctuations of collective nuclear motion. In a recent work,
we addressed this question \cite{R7}. Restricting our treatment at
low energies, we proposed a stochastic mean-field approach for
nuclear dynamics, which incorporates one-body dissipation and
fluctuation mechanisms in accordance with quantal
dissipation-fluctuation theorem. Therefore, the stochastic
mean-field approach provides a powerful microscopic tool for
describing low energy nuclear processes including induced fission,
heavy-ion fusion near barrier energies and spinodal decomposition
of nuclear matter.

Much work has been done to understand the spinodal instabilities
and their connection with liquid gas phase transformation in symmetric
and more recently charge asymmetric nuclear matter. Most of these
investigations have been carried out in the basis of
semi-classical Boltzmann- Langevin (BL) type stochastic transport
models \cite{R8}.  There are two major problems with these
investigations. First of all, numerical simulations of BL model
are not very easy, even with approximate methods, simulations
require large amount of numerical effort. The second problem is
related with the semi-classical description of spinodal
decomposition of nuclear matter. According to our previous
investigations, quantal statistical effects play an important role
in spinodal dynamics \cite{R9,R10}. There are qualitatively
two different regimes during evolution of nuclear collisions in
Fermi energy domain. During the initial regime of heavy
ion-collisions, namely, from touching until formation of hot and
compressed piece of nuclear matter, collisional dissipation and
fluctuations are substantially important. On the other hand,
during expansion of the system into mechanically unstable spinodal
region, collisional effects may be neglected. In the spinodal
region, local density fluctuations, which are accumulated during
the initial regime, are mainly driven by the mean-field until
system breaks up into clusters. Recently proposed stochastic
mean-field approach provides a useful tool for describing spinodal
decomposition of expanding hot piece of nuclear matter. The
approach includes quantum statistical effects and at the same
time, numerical simulations of the approach can be carried out
without much difficulty.

In this work, we study early growth of density fluctuations in
charge asymmetric nuclear matter and investigate quantum
statistical effects on spinodal instabilities and on growth rates of
dominant unstable modes on the basis of stochastic mean-field
approach. In section 2, we present a brief description of the
stochastic mean-field approach. In section 3, we calculate early
growth of density fluctuations, growth rates and phase diagram of
dominant modes in charge asymmetric systems, and study quantal
effects on these quantities. Conclusions are  given in section 4.

\section{STOCHASTIC MEAN-FIELD APPROACH}

In the standard TDHF description of a many-body system,
time-dependent wave function is assumed to be a single Slater
determinant constructed with time-dependent single-particle wave
functions. The standard approach provides a good description for
the average evolution of collective motion, however it severely
restricts fluctuations of collective motion \cite{R1,R2}.
In order to describe fluctuations, we must give up single
determinantal description and consider superposition of
determinantal wave functions. In the stochastic mean-field
description, an ensemble of single-particle density matrices
associated with the ensemble of Slater determinants is generated in
a stochastic framework by retaining only initial correlations
\cite{R7}. A member of single-particle density matrix, indicated
by label $\lambda$ , can be expressed as,
\begin{equation}\label{eq1}
    \rho_{a}^{\lambda}(\vec{r},\vec{r}~',t)=\sum_{ij}\Phi^{\ast}_{i}(\vec{r},t;\lambda)<i|\rho_{a}^{\lambda}(0)
    |j>\Phi_{j}(\vec{r}~',t;\lambda).
\end{equation}
In this expression and in the rest of the paper label $a=n,p$
represents neutron and proton species and
$<i|\rho_{a}^{\lambda}(0)|j>$ are time-independent elements of
density matrix determined by the initial correlations. The main
assumption of the approach is that each matrix element is a
Gaussian random number  specified by a mean value
$\overline{<i|\rho_{a}^{\lambda}(0)|j>}=\delta_{ij}\rho_{a}(i)$
and a variance,
\begin{equation}\label{eq2}
   \overline{<i|\delta\rho_{a}^{\lambda}(0)|j><j'|\delta\rho_{b}^{\lambda}(0)|i'>}=\frac{1}{2}\delta_{ab}\delta_{ii'}\delta_{jj'}
   \{\rho_{a}(i)[1-\rho_{a}(j)]+\rho_{a}(j)[1-\rho_{a}(i)]\}.
\end{equation}
In these expressions $<i|\delta\rho_{a}^{\lambda}(0)|j>$
represents fluctuating elements of initial density matrix,
$\rho_{a}(j)$ denotes the average occupation number.  At zero
temperature, the average occupation numbers are zero and one and
at finite temperature, they are given by the Fermi-Dirac
distribution. In each event, different from the standard TDHF,
time-dependent single-particle wave functions of neutrons and
protons are determined by their own self-consistent mean-field
according to,
\begin{equation}\label{eq3}
    i\hbar\frac{\partial}{\partial t}\Phi_{j}^{a}(\vec{r},t;\lambda)
    =h_{a}^{\lambda}~\Phi_{j}^{a}(\vec{r},t;\lambda).
\end{equation}
Here
$h_{a}^{\lambda}=p^{2}/2m_{a}+U_{a}(n_{n}^{\lambda},n_{p}^{\lambda})$
denotes the self-consistent mean-field Hamiltonian in the event,
which depends on proton and neutron local densities
$n_{a}^{\lambda}(r,t)$. We can express stochastic mean-field
evolution in terms of single-particle density matrices of neutrons
and protons as,
\begin{equation}\label{eq4}
    i\hbar\frac{\partial}{\partial t}\rho_{a}^{\lambda}(t)=[h_{a}^{\lambda},\rho_{a}^{\lambda}(t)].
\end{equation}
In the stochastic mean-field approach an ensemble of
single-particle density matrices is generated associated with
different events. In this approach, we can calculate, not only the
mean value of observables, also probability distribution of
observables. Even if the magnitude of initial fluctuations is
small, in particular in the vicinity of instabilities mean-field
evolution can enhance the fluctuations, and hence events can
substantially deviate from one another.  By projecting on a
collective path, it is demonstrated that the stochastic mean-field
approach incorporates one-body dissipation and one-body
fluctuation mechanisms in accordance with quantal
dissipation-fluctuation relation \cite{R7}.

In this work, we investigate the early growth of density
fluctuations in spinodal region in charge asymmetric nuclear
matter.  For this purpose it is sufficient to consider the linear
response treatment of dynamical evolution \cite{R8}.  The small
amplitude fluctuations of the single-particle density matrix
around an equilibrium state $(\rho_{n}^{0},\rho_{p}^{0})$ are
determined by the linearized TDHF equations. The linearized TDHF
equations for fluctuations of neutron and proton density matrices,
$\delta\rho_{a}^{\lambda}(t)=\rho_{a}^{\lambda}(t)-\rho_{a}^{0}$,
are given by,
\begin{equation}\label{eq5}
    i\hbar\frac{\partial}{\partial t}\delta\rho_{a}^{\lambda}(t)=[h_{a}^{0},\delta\rho_{a}^{\lambda}(t)]
    +[\delta U_{a}^{\lambda}(t),\rho_{a}^{0}].
\end{equation}
Since for infinite matter, the equilibrium state and the
associated mean-field Hamiltonian $h_{a}^{0}$ are homogenous, it
is suitable to analyze these equations in the plane wave
representations,
\begin{eqnarray}\label{eq6}
    &&i\hbar\frac{\partial}{\partial
    t}<\vec{p_{1}}|\delta\rho_{a}(t)|\vec{p_{2}}> = \nonumber \\
    &&[\varepsilon_{a}(\vec{p_{1}})-\varepsilon_{a}(\vec{p_{2}})]
    <\vec{p_{1}}|\delta\rho_{a}(t)|\vec{p_{2}}>
   [\rho_{a}(\vec{p_{1}})-\rho_{a}(\vec{p_{2}})]<\vec{p_{1}}|\delta U_{a}(t)
    |\vec{p_{2}}>.
\end{eqnarray}
According to the basic assumption, matrix elements of the initial
density matrix are Gaussian random numbers. In the plane wave
representation the second moments of the initial correlations is
given by,
\begin{eqnarray}\label{eq7}
    &&\overline{<\vec{p_{1}}|\delta\rho_{a}(0)|\vec{p_{2}}><\vec{p_{2}}'|\delta\rho_{b}(0)|\vec{p_{1}}'>}=\\ \nonumber
    &&\delta_{ab}(2\pi\hbar)^{6}\delta(\vec{p_{1}}-\vec{p_{1}}')
    \delta(\vec{p_{2}}-\vec{p_{2}}')
    \frac{1}{2}[\rho_{a}(\vec{p_{1}})(1-\rho_{a}(\vec{p_{2}}))+
    \rho_{a}(\vec{p_{2}})(1-\rho_{a}(\vec{p_{1}}))],
\end{eqnarray}
where the factor $(2\pi\hbar)^{6}$ arises from normalization of
the plane waves.

\section{GROWTH OF DENSITY FLUCTUATIONS}
\subsection{Spinodal Instabilities}
In this section, we apply the stochastic mean-field approach in
small amplitude limit to investigate spinodal instabilities in
charge asymmetric nuclear matter \cite{R11}.  We note that the
following quantity
\begin{equation}\label{eq8}
    \delta \tilde{n}_{a}(\vec{k},t)=2\int_{-\infty}^{\infty}\frac{d^{3}p}{(2\pi\hbar)^{3}}
    <\vec{p}+\hbar \vec{k}/2|\delta\rho_{a}(t)|\vec{p}-\hbar
    \vec{k}/2>
\end{equation}
defines the Fourier transform of the local density fluctuations of
neutrons and protons. In this expression and in other formulas in
this section, we omit the event label $\lambda$  for clarity of
notation. We can obtain solution of the linear response Eq.
(\ref{eq6}) by employing the standard method of  one-sided Fourier
transform in time,
\begin{equation}\label{eq9}
    \delta \tilde{n}_{a}(\vec{k},\omega)=\int_{0}^{\infty} dt e^{i\omega t}\delta n_{a}(\vec{k},t).
\end{equation}
After transformation, we obtain a set of coupled algebraic equations
for the Fourier transforms of fluctuating parts of local neutron and
proton densities \cite{R12},
\begin{equation}\label{eq10}
    [1+F_{0}^{nn}\chi_{n}(\vec{k},\omega)]\delta
    \tilde{n}_{n}(\vec{k},\omega)+F_{0}^{np}\chi_{n}(\vec{k},\omega)
    \delta \tilde{n}_{p}(\vec{k},\omega)= i A_{n}(\vec{k},\omega)
\end{equation}
and
\begin{equation}\label{eq11}
    [1+F_{0}^{pp}\chi_{p}(\vec{k},\omega)]\delta
    \tilde{n}_{p}(\vec{k},\omega)+F_{0}^{pn}\chi_{p}(\vec{k},\omega)
    \delta \tilde{n}_{n}(\vec{k},\omega)= i A_{p}(\vec{k},\omega).
\end{equation}
In these expressions, derivative of the mean-field potential
$U_{a}(n_{n},n_{p})$  evaluated at the equilibrium density
$F_{0}^{ab}=\left(\partial U_{b}/\partial n_{a}\right)_{0}$ denotes
the zero-order Landau parameters and $\chi_{a}(\vec{k},\omega)$  is
the Lindhard functions associated with neutron and proton
distributions,
\begin{equation}\label{eq12}
    \chi_{a}(\vec{k},\omega)=-2\int_{-\infty}^{\infty}\frac{d^{3}p}{(2\pi\hbar)^{3}}
    \frac{\rho_{a}(\vec{p}-\hbar \vec{k}/2)-\rho_{a}(\vec{p}
    +\hbar \vec{k}/2)}
    {\hbar\omega-\vec{p}\cdot\hbar\vec{k}/m}.
\end{equation}
The source terms $ A_{a}(\vec{k},\omega)$  are determined by the
initial conditions,
\begin{equation}\label{eq13}
     A_{a}(\vec{k},\omega)=2\hbar\int_{-\infty}^{\infty}\frac{d^{3}p}{(2\pi\hbar)^{3}}
    \frac{<\vec{p}+\hbar \vec{k}/2|\delta\rho_{a}(0)|\vec{p}-\hbar \vec{k}/2>}
    {\hbar\omega-\vec{p}\cdot\hbar\vec{k}/m}.
\end{equation}
The solution of the coupled algebraic equations for Fourier
transform of density fluctuations is given by,
\begin{equation}\label{eq14}
    \delta \tilde{n}_{n}(\vec{k},\omega)=i\frac{[1+F_{0}^{pp}\chi_{p}(\vec{k},\omega)]
    A_{n}(\vec{k},\omega)-F_{0}^{np}\chi_{n}(\vec{k},\omega)A_{p}(\vec{k},\omega)}
    {\varepsilon(\vec{k},\omega)}
\end{equation}
 and
\begin{equation}\label{eq15}
    \delta \tilde{n}_{p}(\vec{k},\omega)=i\frac{[1+F_{0}^{nn}\chi_{n}(\vec{k},\omega)]
    A_{p}(\vec{k},\omega)-F_{0}^{pn}\chi_{p}(\vec{k},\omega)A_{n}(\vec{k},\omega)}
    {\varepsilon(\vec{k},\omega)},
\end{equation}
where the quantity
\begin{equation}\label{eq16}
    \varepsilon(\vec{k},\omega)=1+F_{0}^{nn}\chi_{n}(\vec{k},\omega)+
    F_{0}^{pp}\chi_{p}(\vec{k},\omega)+[F_{0}^{nn}F_{0}^{pp}-F_{0}^{np}F_{0}^{pn}]
    \chi_{n}(\vec{k},\omega)\chi_{p}(\vec{k},\omega)
\end{equation}
denotes the susceptibility.

Time dependence of Fourier transform of density fluctuations
$\delta\tilde{n}_{a}(\vec{k},t)$ is determined by taking the inverse
transformation of Eqs. (\ref{eq14}) and (\ref{eq15}) \cite{R13}. The
inverse Fourier transformations in time can be calculated with the
help of residue theorem, keeping only the growing and decaying
collective poles we find,
\begin{equation}\label{eq17}
    \delta \tilde{n}_{a}(\vec{k},t)=\delta
    n_{a}^{+}(\vec{k})e^{+\Gamma_{k}t}+
    \delta n_{a}^{-}(\vec{k})e^{-\Gamma_{k}t},
\end{equation}
 where the initial amplitude of density fluctuations   are given by
 \begin{equation}\label{eq18}
    \delta n_{n}^{\mp}(\vec{k})=-\left\{\frac{[1+F_{0}^{pp}\chi_{p}(\vec{k},\omega)]
    A_{n}(\vec{k},\omega)-F_{0}^{np}\chi_{n}(\vec{k},\omega)A_{p}(\vec{k},\omega)}
    {\partial \varepsilon(\vec{k},\omega)/\partial\omega} \right\}_{\omega={\mp}i\Gamma_{k}}
\end{equation}
and
\begin{equation}\label{eq19}
    \delta n_{p}^{\mp}(\vec{k})=-\left\{\frac{[1+F_{0}^{nn}\chi_{n}(\vec{k},\omega)]
    A_{p}(\vec{k},\omega)-F_{0}^{pn}\chi_{p}(\vec{k},\omega)A_{n}(\vec{k},\omega)}
    {\partial \varepsilon(\vec{k},\omega)/\partial\omega} \right\}_{\omega={\mp}i\Gamma_{k}}.
\end{equation}
Growth and decay rates $\omega={\mp}i\Gamma_{k}$ are determined from
the dispersion relation $\varepsilon(\vec{k},\omega)=0$, i.e. from
the roots of susceptibility.

In numerical calculations we employ the same effective Skyrme
potential as in reference \cite{R11},
\begin{equation}\label{eq20}
    U_{a}(n_{n},n_{p})=A\left(\frac{n}{n_{0}}\right)+B\left(\frac{n}{n_{0}}\right)^{\alpha+1}+
    C\left(\frac{n'}{n_{0}}\right)\tau_{a}+\frac{1}{2}\frac{dC}{dn}\frac{n'^{2}}{n_{0}}-D\Delta
    n+D'\Delta n'\tau_{a}
\end{equation}
where $n=n_{n}+n_{p}$ and $n'=n_{n}-n_{p}$ are total and relative
densities, and $\tau_{a}=+1$ for neutrons and $\tau_{a}=-1$  for
protons. The parameters $A=-356.8 ~ MeV$, $B=+303.9 ~ MeV$,
$\alpha=1/6$ and $D=+130.0 ~ MeVfm^{5}$ are adjusted to reproduce
the saturation properties of symmetric nuclear matter: The binding
energy  $\varepsilon_{0}=15.7 ~ MeV/nucleon$ and zero pressure at
the saturation density $n_{0}=0.16 ~ fm^{-3}$, compressibility
modulus $K=201 ~ MeV$ and the surface energy coefficient in the
Weizsacker mass formula $a_{surf}=18.6 ~ MeV$ \cite{R14}. Magnitude
of the parameter $D'=+34 ~ MeVfm^{5}$ is close to magnitude given by
the $SkM^{*}$ interaction \cite{R15}. The potential symmetry energy
coefficient is $C(n)=C_{1}-C_{2}(n/n_{0})^{\alpha}$ with
$C_{1}=+124.9 ~ MeV$ and $C_{2} = 93.5 ~ MeV$. These parameters for
the symmetry energy coefficient in Weizsacker mass formula, at
saturation density gives
$a_{sym}=\varepsilon_{F}(n_{0})/3+C(n_{0})/2=36.9/3+31.4/2=28.0 ~
MeV$.

\begin{figure}
 \includegraphics[width=12cm]{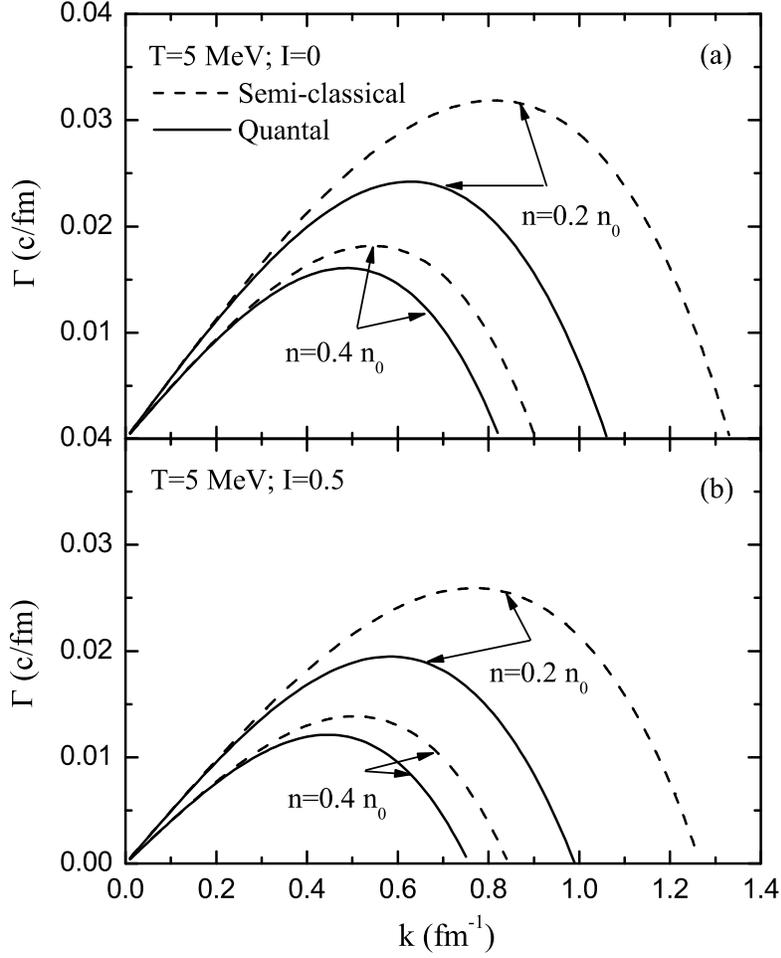}
 \caption{\label{fig1}Growth rates of unstable modes as a function of wave number
 in spinodal region corresponding initial densities   and   at a temperature $T=5~MeV$.
(a) for initial asymmetry $I=0.0$ , (b)  for initial asymmetry
$I=0.5$.}
\end{figure}

\begin{figure}
 \includegraphics[width=12cm]{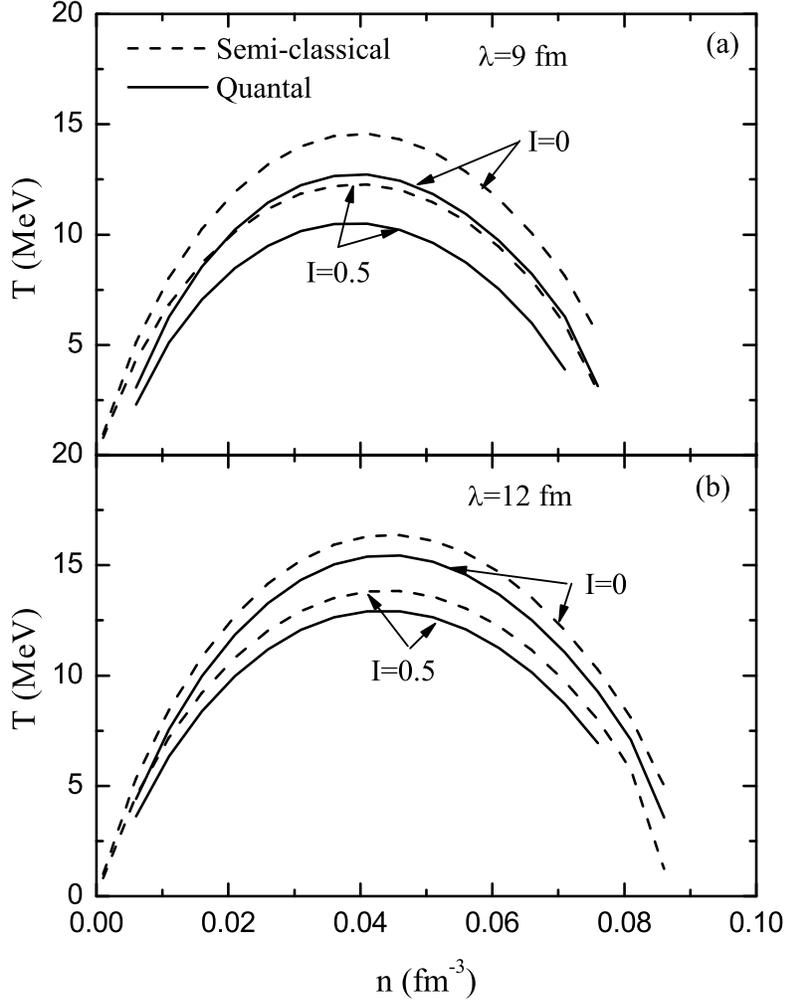}
 \caption{\label{fig2}Boundary of spinodal region in density-temperature
  plane corresponding to initial charge asymmetries $I=0.0$  and  $I=0.5$  for the
  unstable mode: (a) with wavelength $\lambda=9 ~ fm$ , (b) with wavelength $\lambda=12 ~ fm$.}
\end{figure}

As an example, Fig. \ref{fig1}(a) shows the growth rates of unstable
modes as a function of wave number in the spinodal region
corresponding to initial density $n=0.2~n_{0}$ and $n=0.4~n_{0}$ for
initial asymmetry $I=0.0$ at a temperature $T=5 ~ MeV$. The initial
charge asymmetry is defined according to
$I=(n_{n}^{0}-n_{p}^{0})/(n_{n}^{0}+n_{p}^{0})$. In this figure and
also in other figures, solid-lines and dashed-lines show quantal and
semi-classical results, respectively. Since, at low densities, wave
numbers of most unstable modes are comparable to Fermi momentum,
long-wavelength expansion of the Linhard function is not valid, and
hence there is important quantal effect in the dispersion relation.
At the initial density $n=0.2~n_{0}$ and the initial asymmetry
$I=0.0$, in the quantal calculations unstable modes are confined to
a narrower range centered around wavelengths $\lambda \approx 8-10
fm$, as compared to a broader range centered around $\lambda \approx
7 fm$ in the semi-classical calculations. Growth rates in
semi-classical framework are determined by the roots of
semi-classical susceptibility, which is defined as in Eq.
(\ref{eq16}) by taking the Lindhard functions
$\chi_{a}(\vec{k},\omega)$ in the long wavelength limit given by Eq.
(\ref{eq28}). As a result, in the quantal calculations, the source
has a tendency to break up into larger fragments as compared to the
semi-classical calculations. Also, due to quantum effects, the
maximum of dispersion relation is reduced by about a factor $3/4$.
Therefore, fluctuations take more time to develop when quantum
effects are introduced. At higher initial density $n=0.4~n_{0}$ , in
both quantal and semi-classical calculations, dispersion relation is
shifted towards longer wavelengths and it exhibits a similar trend
as the one at the initial density $n=0.2~n_{0}$. This quantal effect
in dispersion relation of unstable modes was pointed out in the case
of symmetric matter in a previous publication \cite{R16}. Charge
asymmetric nuclear matter exhibits a similar behavior as seen from
figure \ref{fig1}(b), which shows dispersion relation corresponding
to initial densities $n=0.2~n_{0}$ and $n=0.4~n_{0}$  for initial
charge asymmetry $I=0.5$ at a temperature $T=5 ~ MeV$. Figs.
\ref{fig2}(a) and \ref{fig2}(b) shows the boundary of spinodal
region in density-temperature plane corresponding to initial charge
asymmetries $I=0.0$ and $I=0.5$  at a temperature  $T=5 ~ MeV$ for
the unstable modes with wavelengths $\lambda=9 ~fm$  and
$\lambda=12~ fm$, respectively. It is seen that with increasing
charge asymmetry, spinodal region shrinks to smaller size in both
quantal and semi-classical calculations. Furthermore, unstable modes
are quite suppressed by quantal effects as compared to the
semi-classical results in both symmetric and asymmetric matter.
Results of semi-classical calculation are in agreement with the
result obtained in reference \cite{R11}.

\subsection{Growth of Density fluctuations}

In this section, we calculate early growth of local density
fluctuations in charge asymmetric nuclear matter. Spectral intensity
of density correlation function $\tilde{\sigma}_{ab}(\vec{k},t)$ is
related to the second moment of Fourier transform of density
fluctuations according to,
\begin{equation}\label{eq21}
    \tilde{\sigma}_{ab}(\vec{k},t)(2\pi)^{3}\delta(\vec{k}-\vec{k}~')=
    \overline{\delta\tilde{n}_{a}(\vec{k},t)\delta\tilde{n}_{b}(-\vec{k}~',t)}.
\end{equation}
We calculate the spectral functions using the solution (\ref{eq17})
and employing expression (\ref{eq7}) for the initial correlations to
find,
\begin{equation}\label{eq22}
    \tilde{\sigma}_{ab}(\vec{k},t)=
    \frac{E^{+}_{ab}(\vec{k},i\Gamma_k)}{|[\partial\varepsilon(\vec{k},\omega)
    /\partial\omega]_{\omega=i\Gamma_k}|^{2}}(e^{2\Gamma_{k}t}+e^{-2\Gamma_{k}t})+
     \frac{2E^{-}_{ab}(\vec{k},i\Gamma_k)}{|[\partial\varepsilon(\vec{k},\omega)
    /\partial\omega]_{\omega=i\Gamma_k}|^{2}}
\end{equation}
where quantities $E^{\mp}_{ab}(\vec{k},i\Gamma_k)$, $a,b=n,p$, are
given by,
\begin{equation}\label{eq23}
    E^{\mp}_{nn}(\vec{k},i\Gamma_k)=4\hbar^{2}(1+F^{pp}_{0}\chi_{p})^{2}I^{\mp}_{n}+
    4\hbar^{2}(F^{np}_{0}\chi_{n})^{2}~I^{\mp}_{p}~,
\end{equation}
\begin{equation}\label{eq24}
    E^{\mp}_{pp}(\vec{k},i\Gamma_k)=4\hbar^{2}(1+F^{nn}_{0}\chi_{n})^{2}I^{\mp}_{p}+
    4\hbar^{2}(F^{pn}_{0}\chi_{p})^{2}~I^{\mp}_{n}
\end{equation}
and
\begin{equation}\label{eq25}
    E^{\mp}_{np}(\vec{k},i\Gamma_k)=-4\hbar^{2}(1+F^{pp}_{0}\chi_{p})F^{pn}_{0}\chi_{p}~I^{\mp}_{n}-4\hbar^{2}
    (1+F^{nn}_{0}\chi_{n})F^{np}_{0}\chi_{n}~I^{\mp}_{p}
\end{equation}
with
\begin{equation}\label{eq26}
    I^{\mp}_{a}=\int\frac{d^{3}p}{(2\pi\hbar)^{3}}\frac{(\hbar\Gamma_k)^{2}\mp(\vec{p}\cdot\hbar\vec{k}/m)^{2}}
    {[(\hbar\Gamma_k)^{2}+(\vec{p}\cdot\hbar\vec{k}/m)^{2}]^{2}}
    \rho_{a}(\vec{p}+\hbar\vec{k}/2)[1-\rho_{a}(\vec{p}-\hbar\vec{k}/2)].
\end{equation}
Semi-classical limit of these expressions are obtained by replacing
the integrals $I^{\mp}_{a}$  and $\chi_{a}(\vec{k},\omega)$ with
following expressions in the long wave-length limit,
\begin{equation}\label{eq27}
    I^{\mp}_{a}(sc)=\int\frac{d^{3}p}{(2\pi\hbar)^{3}}\frac{(\hbar\Gamma_k)^{2}\mp(\vec{p}\cdot\hbar\vec{k}/m)^{2}}
    {[(\hbar\Gamma_k)^{2}+(\vec{p}\cdot\hbar\vec{k}/m)^{2}]^{2}}
    \rho_{a}(\vec{p})[1-\rho_{a}(\vec{p}]
\end{equation}
and
\begin{equation}\label{eq28}
    \chi_{a}^{sc}(\vec{k},\omega)=-2\int_{-\infty}^{\infty}\frac{d^{3}p}{(2\pi\hbar)^{3}}
    \frac{(\vec{p}\cdot\hbar\vec{k}/m)^{2}}{(\hbar\Gamma_k)^{2}+(\vec{p}\cdot\hbar\vec{k}/m)^{2}}
    \frac{\partial}{\partial\varepsilon}\rho_{a}.
\end{equation}

\begin{figure}[htb]
\includegraphics[width=12cm,height=15cm]{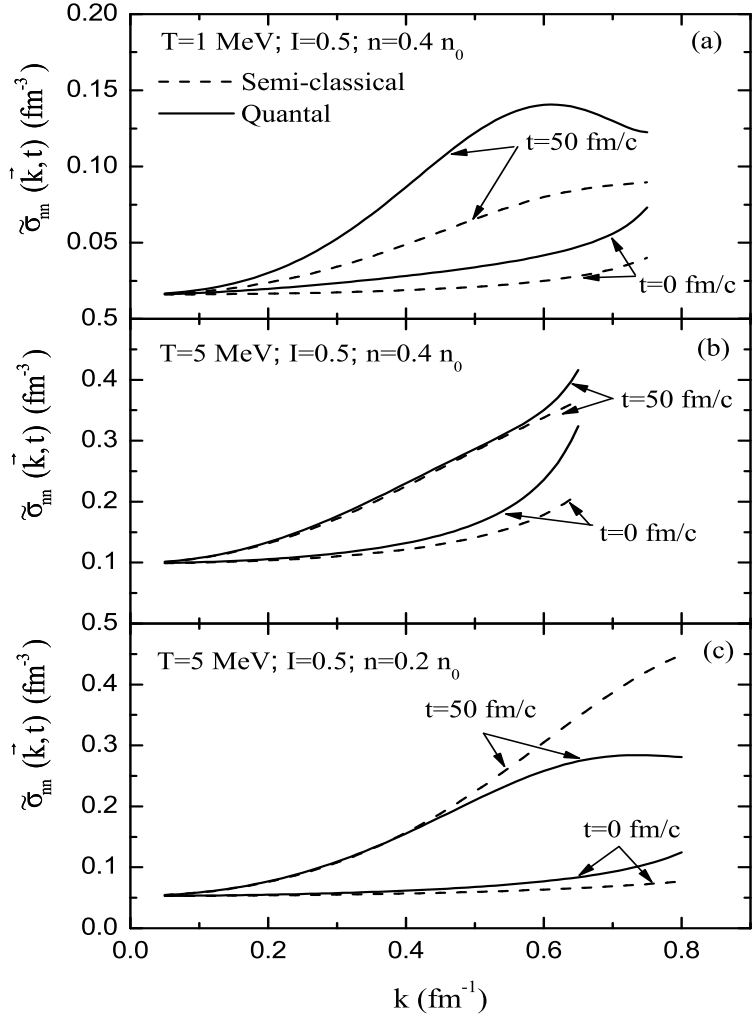}
\caption{\label{fig3}Spectral intensity
$\tilde{\sigma}_{nn}(\vec{k},t)$ of neutron-neutron density
correlation function as function of wave number k at times $t=0$ and
$t=50 ~ fm/c$ for the initial charge asymmetry $I=0.5$ : (a) for
density $n=0.4~n_{0}$ at temperature $T=1 ~ MeV$, (b) for density
$n=0.4~n_{0}$ at temperature  $T=5 ~ MeV$ , (c)  for density
$n=0.2~n_{0}$ at temperature $T=5 ~ MeV$.}
\end{figure}

Figs. \ref{fig3}(a) and \ref{fig3}(b) shows spectral intensity
$\tilde{\sigma}_{nn}(\vec{k},t)$ of neutron-neutron density
correlation function as function of wave number at times $t=0$ and
$t= 50 ~ fm/c$ for density  $n=0.4~n_{0}$ and the initial charge
asymmetry $I=0.5$ at temperature  $T=1 ~ MeV$ and $T=5 ~ MeV$,
respectively.

\begin{figure}[htb]
\includegraphics[width=12cm,height=15cm]{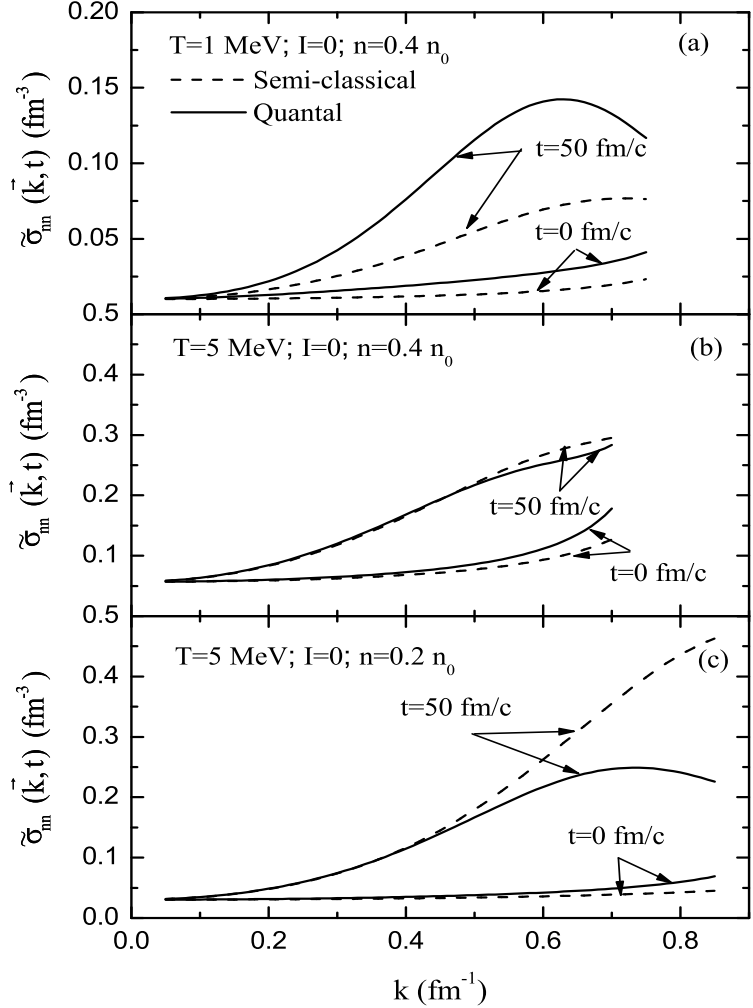} \caption{\label{fig4} Same as Fig. 3 but for asymmetry
$I=0.0$.}
\end{figure}

As mentioned above, in all figures solid-lines and dashed-lines
indicate quantal and semi-classical results, respectively. As seen,
in particular at towards the high end of the wave number spectrum,
considerable quantal effects are present at initial fluctuations.
Quantum statistical effects in the initial fluctuations become even
larger at smaller temperatures. In fact at zero temperature, since
the quantities $I^{\mp}_{a}(sc)$ becomes zero, spectral functions
vanish $\tilde{\sigma}_{ab}(\vec{k},t)=0$. However, in quantal
calculations spectral functions remains finite even at zero
temperature, reflecting quantum zero point fluctuations of the local
density. Looking at the results at $t=50 ~ fm/c$ , we observe that
largest growth occurs over the range of wave numbers corresponding
to the range of dominant unstable modes. At $T=5 ~ MeV$, magnitude
of fluctuations is about the same in both quantal and semi-classical
calculations. At the lower temperature  $T=1 ~MeV$, magnitude of
fluctuations in the most unstable range is nearly doubled in quantal
calculations as compared to semi-classical calculations. Fig.
\ref{fig3}(c) shows spectral intensity
$\tilde{\sigma}_{nn}(\vec{k},t)$ as function of wave number at times
$t=0$  and $t=50 ~ fm/c$ at a lower density $n=0.2~n_{0}$ for
initial charge asymmetry $I=0.5$ and temperature $T=5 ~MeV$. At the
lower density, growth rates of dominant modes in the semi-classical
limit are considerably larger than those of quantal calculations.
Consequently, the result of semi-classical calculations at time
$t=50 ~ fm/c$ overshoots the result of quantal calculations over the
range of dominant modes. Fig. \ref{fig4} illustrates that the
spectral intensity for symmetric matter has similar properties as
for asymmetric matter with $I=0.0$.

We note that quantal effects enter into the spectral density in two
different ways: (i) quantal effects in growth rates of modes and
(ii) quantum statistical effects on the initial density
fluctuations, which becomes increasingly more important at lower
temperatures. We  also note that in determining time evolution of
$\delta \tilde{n}(\vec{k},t)$ with the help of residue theorem,
there are other contributions arising from non-collective poles of
susceptibility $\varepsilon(\vec{k},\omega)$ and from poles of
$A_{a}(\vec{k},\omega)$. These contributions, in particular towards
short wavelengths,  are important at the initial state, however they
damp out in a short time interval \cite{R17}. Therefore the
approximate expression (\ref{eq22}) for the spectral intensity
$\tilde{\sigma}(\vec{k},t)$  of density fluctuations becomes more
accurate for increasing time.

\begin{figure}[htb]
\includegraphics[width=12cm,height=15cm]{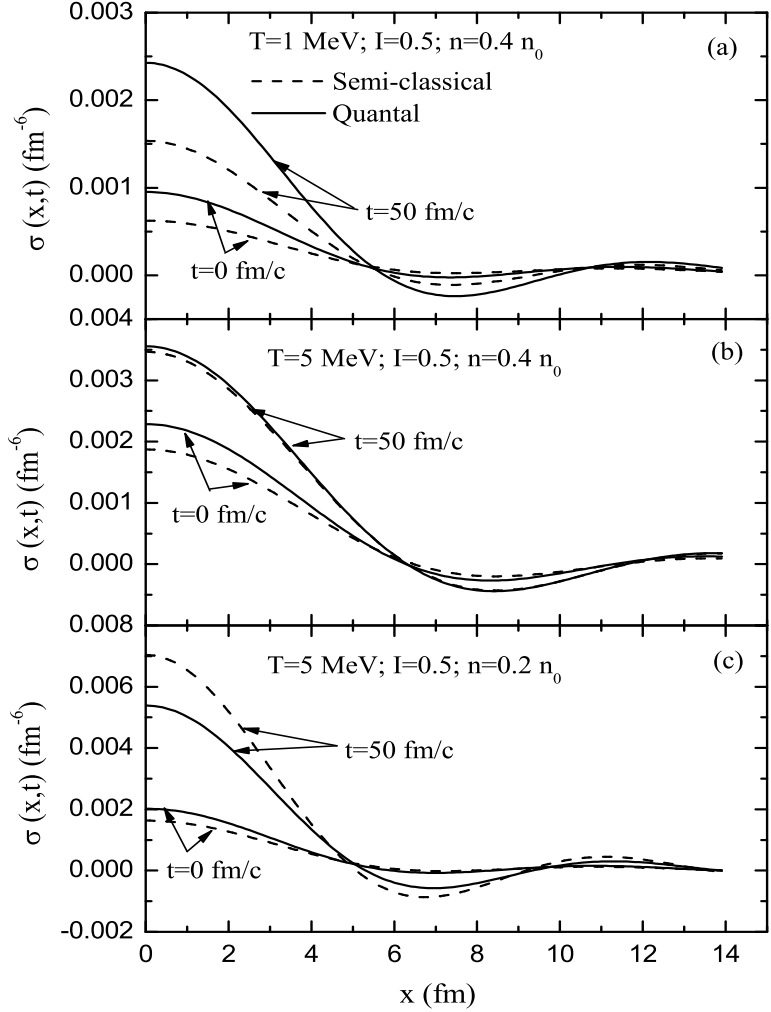}
\caption{\label{fig5} Density correlation function $\sigma(x,t)$
as a function of distance
$x=\mid\vec{r}-\vec{r~'}\mid$ between two space points at times $t=0$ and
$t=50 ~ fm/c$ and the initial charge asymmetry $I=0.5$ : (a) for
density $n=0.4~n_{0}$ at temperature $T=1 ~MeV$, (b) for density
$n=0.4~n_{0}$ at temperature $T=5 ~MeV$, (c) for density
$n=0.2~n_{0}$  at temperature $T=5 ~MeV$.}
\end{figure}

\begin{figure}[htb]
\includegraphics[width=12cm,height=15cm]{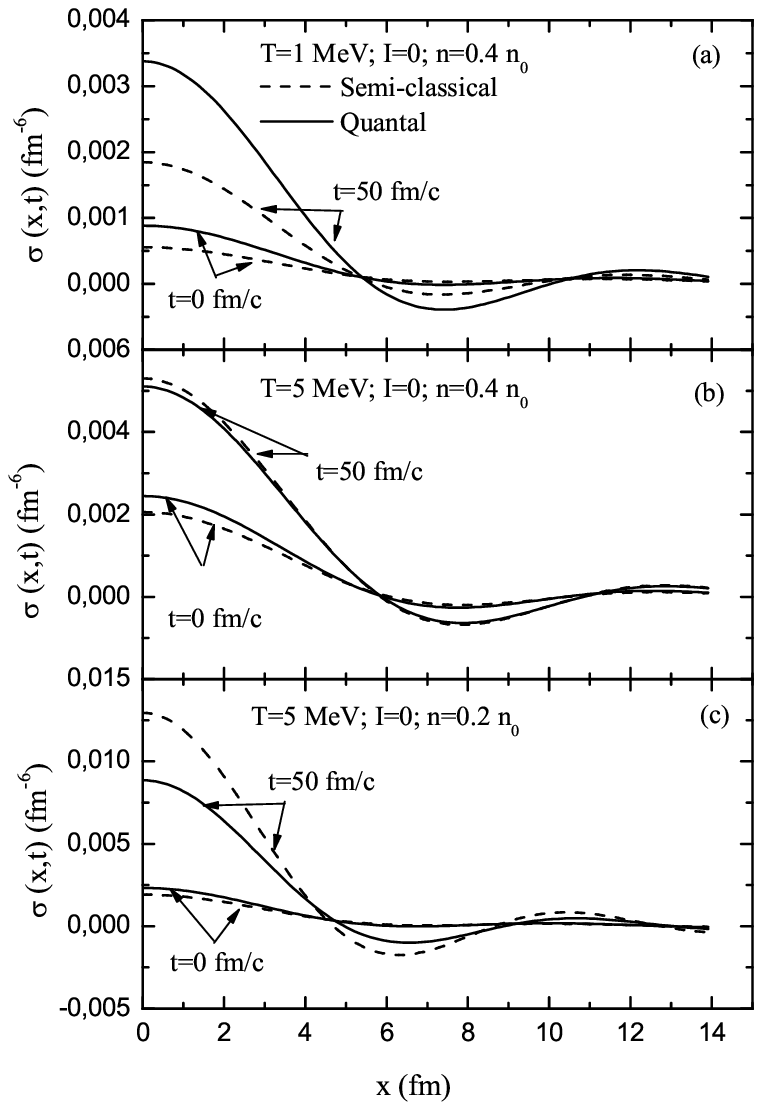}
\caption{\label{fig6} Same as Fig. 5 but for asymmetry $I=0.0$.}
\end{figure}

Local density fluctuations $\delta n_{a}(\vec{r},t)$ are determined
by the Fourier transform of $\delta \tilde{n}_{a}(\vec{k},t)$. In
terms of spectral intensity $\tilde{\sigma}_{ab}(\vec{k},t)$, which
is defined in Eq. (\ref{eq21}), equal time density correlation
function as a function of distance between two space locations is
expressed as,
\begin{equation}\label{eq29}
    \sigma_{ab}(|\vec{r}-\vec{r}~'|,t)=\overline{\delta
    n_{a}(\vec{r},t)\delta
    n_{b}(\vec{r}~',t)}=\int\frac{d^{3}k}{(2\pi)^{3}}e^{i\vec{k}
    \cdot(\vec{r}-\vec{r}~')}\tilde{\sigma}_{ab}(\vec{k},t).
\end{equation}
Total density correlation function is given by sum over neutrons and
protons and cross-term, $\sigma(|\vec{r}-\vec{r}~'|,t)=
\sigma_{nn}(|\vec{r}-\vec{r}~'|,t)+\sigma_{pp}(|\vec{r}-\vec{r}~'|,t)
+2\sigma_{np}(|\vec{r}-\vec{r}~'|,t)$. The behavior of density
correlation function as a function of initial density and
temperature carries valuable information about the unstable dynamics
of the matter in the spinodal region. As an example,   Figs.
\ref{fig5}(a) and \ref{fig5}(b)  illustrate total density
correlation function as a function of distance between two space
points at times $t=0$ and $t=50 ~ fm/c$ at density $n=0.4~n_{0}$ and
the initial charge asymmetry $I=0.5$ for temperatures $T=1 ~MeV$ and
$T=5 ~MeV$, respectively. At temperature $T=5 ~ MeV$, quantal
effects are not important, and hence semi-classical calculations
provide good approximation for density correlation function.
However, at lower temperature $T=1 ~MeV$, semi-classical
calculations severely underestimates peak value of density
correlation function.  Fig. \ref{fig5}(c) shows density correlation
function at times $t=0$ and $t=50 ~~fm/c$ at a lower density
$n=0.2~n_{0}$ for initial charge asymmetry  $I=0.5$ and a
temperature $T=5 ~ MeV$. On the other hand, at lower density,
semi-classical approximation overestimates the peak value of the
correlation function. As indicated above, this is due to the fact
that growth rates of dominant modes in semi-classical limit are
considerably larger than those obtained in quantal calculations. For
asymmetry $I=0.0$, as seen from Fig. \ref{fig6}, behavior of density
correlation function is similar to the charge asymmetric case.
Complementary to the dispersion relation, correlation length of
density fluctuations provides an additional measure for the average
size of primary fragmentation pattern. We can estimate the
correlation length of density fluctuations as the width of
correlation function at half maximum. Correlation length depends on
density, and to some extend, depends on temperature as well. From
these figures, we can estimate that the correlation length of
density fluctuations is about $3.5 ~ fm$  at density $n=0.4~n_{0}$,
and about $3.0 ~ fm$ at density $n=0.2~n_{0}$.

\begin{figure}[htb]
\includegraphics[width=12cm]{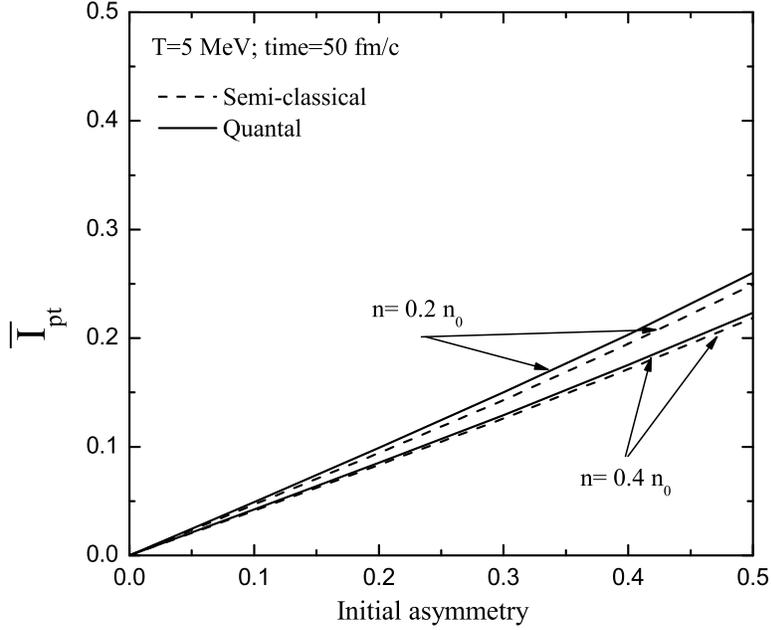}
\caption{\label{fig7} Perturbation asymmetry as function of initial
asymmetry at temperature $T=5 ~ MeV$  for densities $n=0.2~n_{0}$
and  $n=0.4~n_{0}$.}
\end{figure}

During spinodal decomposition, initial charge asymmetry shifts
towards symmetry in liquid phase while gas phase moves toward
further asymmetry. As a result, produced fragments are more
symmetric than the charge asymmetry of the source. This interesting
fact is experimentally observed and it may provide a useful guidance
to gain information about symmetry energy in low density nuclear
matter. For each event, we can define perturbation charge asymmetry
during early evolution of density fluctuations as,
\begin{equation}\label{eq30}
    I_{pt}=\frac{\delta n_{n}(\vec{r},t)-\delta n_{p}(\vec{r},t)}
    {\delta n_{n}(\vec{r},t)+\delta
    n_{p}(\vec{r},t)}=
    \frac{[\delta n_{n}(\vec{r},t)]^{2}-[\delta n_{p}(\vec{r},t)]^{2}}
    {[\delta n_{n}(\vec{r},t)+\delta n_{p}(\vec{r},t)]^{2}}.
\end{equation}
We are interested in the ensemble average value of this quantity,
which can approximately be evaluated according to
\begin{equation}\label{eq31}
    \overline{I}_{pt}\approx\frac{\sigma_{nn}(t)-\sigma_{pp}(t)}{\sigma_{nn}(t)+
    2\sigma_{np}(t)+\sigma_{pp}(t)}.
\end{equation}
where $\sigma_{ab}(t)=\sigma_{ab}(\mid\vec{r}-\vec{r~'}\mid=0,t)$.
The average value of the
perturbation asymmetry is nearly independent of time. As an example,
Fig. \ref{fig7} shows this quantity as function of initial asymmetry
at temperature  $T=5 ~ MeV$ for densities $n=0.2~n_{0}$ and
$n=0.4~n_{0}$.  As a result of the driving force of symmetry energy,
perturbation asymmetry drifts towards symmetry. At this temperature
quantal effects do not play an important role and these calculations
are consistent with results of ref. \cite{R11}.

\section{CONCLUSIONS} Recently proposed stochastic mean-field
theory incorporates both one-body dissipation and fluctuation
mechanisms in a manner consistent with quantal
fluctuation-dissipation theorem of non-equilibrium statistical
mechanics. Therefore, this approach provides a powerful tool for
microscopic description of low energy nuclear processes in which
two-body dissipation and fluctuation mechanisms do not play
important role. The low energy processes include induced fission,
heavy-ion fusion near barrier energies, spinodal decomposition of
nuclear matter and nuclear multi-fragmentations. In this work we
investigate quantal effects on spinodal instabilities and early
growth of density fluctuations in charge asymmetric nuclear matter.
For this purpose it is sufficient to consider the linear response
treatment of the stochastic mean-field approach. Retaining only
growing and decaying collective modes, it is possible to calculate
time evolution of spectral intensity of density correlation function
and the density correlation function itself including quantum
statistical effects. Growth rates of unstable collective modes are
determined from a quantal dispersion relation, i.e. from the roots
of susceptibility.  Due to quantal effects, growth rates of unstable
modes, in particular with wave numbers larger than the Fermi
momentum, are strongly suppressed. As a result, dominant collective
modes are shifted to longer wavelengths than those obtained in the
semi-classical description with the same effective interaction, in
both symmetric and asymmetric matter. The size of spinodal zone
associated with these modes is reduced by the quantal effects. In
calculation of density correlation function, quantal effects enter
into the description through the growth rates of the modes and
through the initial density fluctuations. Quantum statistical
influence on density correlation functions grows larger at lower
temperatures and also at lower densities. Quantal effects appear to
be important for a quantitative description of spinodal
instabilities and growth of density fluctuations in an expanding
nuclear system. Stochastic mean-field approach incorporates both
one-body dissipation and fluctuations mechanisms in a manner
consistent with dissipation-fluctuation theorem. Therefore, it will
be very interesting to investigate spinodal decomposition of an
expanding nuclear system in this framework. We also note that
numerical effort in
simulation of stochastic mean-field approach is not so much greater
than the effort required in solving ordinary three dimensional time
dependent Hartree-Fock equations.

\begin{acknowledgments}
One of us (S.A.) gratefully acknowledges TUBITAK for a partial
support and the Physics Department of Middle East Technical
University for warm hospitality extended to him during his visit.
This work is supported in part by the US DOE grant No.
DE-FG05-89ER40530 and  in part by the TUBITAK grant No. 107T691.
\end{acknowledgments}

\end{document}